# Ideas by S.V. Vonsovsky and Modern Model Treatment of Magnetism

V. Yu. Irkhin

Institute of Metal Physics, 620041 Ekaterinburg, Russia

Valentin.Irkhin@imp.uran.ru

**Keywords:** magnetism, many-electron models, Kondo effect

**Abstract.** A review of fundamental works by Shubin and Vonsovsky on the formulation of the polar and *s–d(f)* exchange models is given. Their ideas are compared with subsequent developments in the theory of magnetism in *d*- and *f*-metals and their compounds. Modern approaches including various slave-boson and slave-fermion representations, formation of exotic quasiparticles etc. are discussed. Internal connections between different many-electron models (the Heisenberg, Hubbard, *t–J*, Anderson Hamiltonians) are presented. Description of anomalous rare-earth and actinide compounds (Kondo lattices, systems with heavy fermions and non-Fermi-liquid behavior) within the framework of the *s–d(f)* exchange model and related models is considered.

**Introduction**

The papers by S.P. Shubin and S.V. Vonsovsky on the polar model [1] put forward a program of building up a systematic theory of solids with account of electron correlations. Such a theory should explain simultaneously electric and magnetic properties of metals and combine localized and itinerant features of *d*-electrons. This program was extensively developed by many investigators, but is not fully completed up to now.

The *s–d(f)* exchange model [2] provided a basis to describe transport properties of transition 3*d*-metals and magnetism of 4*f*-metals. Later this model was applied to explain electronic properties of anomalous rare-earth and actinides compounds, including Kondo lattices and heavy-fermion systems.

The present paper is devoted to evolution of the ideas of many-electron models developed in the works by S.V. Vonsovsky and his colleagues.

**Polar model**

The many-electron polar model of a crystal [1] was proposed as a synthesis of the homeopolar Heisenberg model describing a localized-moment system and the Slater determinant approach treating many-electron system of a metal. The initial formulation of the model included the electron hopping and all the types of electron-electron interaction. The corresponding Hamiltonian in the second quantization representation was written down by Bogolyubov:

$$H = \sum_{\nu_1 \neq \nu_2, \sigma} t_{\nu_1 \nu_2} c^\dagger_{\nu_1 \sigma} c_{\nu_2 \sigma} + \frac{1}{2} \sum_{\nu_i \sigma_1 \sigma_2} I_{\nu_1 \nu_2 \nu_3 \nu_4} c^\dagger_{\nu_1 \sigma_1} c^\dagger_{\nu_2 \sigma_2} c_{\nu_4 \sigma_2} c_{\nu_3 \sigma_1}.$$

Later an important step was made by Hubbard [3] who has picked up the most important part of the Coulomb interaction, the on-site repulsion $U = I(\nu\nu\nu\nu)$, to obtain

$$H = \sum_{\mathbf{k}\sigma} t_\mathbf{k} c^\dagger_{\mathbf{k}\sigma} c_{\mathbf{k}\sigma} + U \sum_i c^\dagger_{i\uparrow} c_{i\uparrow} c^\dagger_{i\downarrow} c_{i\downarrow},$$

with $t_\mathbf{k}$ being the band spectrum. The Hubbard model enables one to obtain a formal interpolation between the atomic and band limits. It contains a rich physics and is widely used to describe various phenomena: ferromagnetism, metal–insulator transition etc. At the same time, a strict treatment of this simple model is a very difficult problem.

In the case of strong correlations, it is useful to use the atomic representation which takes into account the intraatomic correlations exactly [4,5]. This was firstly made by Hubbard [4] who introduced generalized projection $X$-operators
$$X(\Gamma,\Gamma') = |\Gamma\rangle\langle\Gamma'|$$
where $|\Gamma\rangle$ are the exact eigenstates of the atomic Hamiltonian. For example, in the simplest case of $s$-electrons we have $\Gamma = 0$, $\sigma = \pm (\uparrow, \downarrow)$, 2, with $|0\rangle$ being the empty state (hole) and $|2\rangle$ doubly-occupied singlet state (double) on a site. Then we derive
$$X(0,0) = (1-\hat{n}_\uparrow)(1-\hat{n}_\downarrow), \quad X(2,2) = \hat{n}_\uparrow \hat{n}_\downarrow,$$
$$X(\sigma,\sigma) = \hat{n}_\sigma(1-\hat{n}_{-\sigma}), \quad X(\sigma,-\sigma) = a_\sigma^\dagger a_{-\sigma},$$
$$X(\sigma,0) = a_\sigma^\dagger(1-\hat{n}_{-\sigma}), \quad X(2,\sigma) = -\sigma a_{-\sigma}^\dagger \hat{n}_\sigma.$$
General commutation relations for $X$-operators at sites $v$ and $v'$ read
$$[X_v(\Gamma,\Gamma'), X_{v'}(\Gamma'',\Gamma''')]_\pm = \delta_{vv'}\{X(\Gamma,\Gamma''')\delta_{\Gamma'\Gamma''} \pm X(\Gamma'',\Gamma')\delta_{\Gamma\Gamma'''}\}$$
where the plus sign corresponds to the case where both $X$-operators have the Fermi type, i.e. change the number of electrons by an odd number, and minus sign to all the other cases. The one-band Hamiltonian of a crystal in the many-electron (ME) representation reads
$$H = U\sum_v X_v(2,2) + \sum_{v_1 v_2 \sigma} \{t_{v_1 v_2}^{(00)} X_{v_1}(\sigma,0) X_{v_2}(0,\sigma) + t_{v_1 v_2}^{(22)} X_{v_1}(2,\sigma) X_{v_2}(\sigma,2) +$$
$$+ \sigma t_{v_1 v_2}^{(02)} [X_{v_1}(\sigma,0) X_{v_2}(-\sigma,2) + X_{v_1}(2,-\sigma) X_{v_2}(0,\sigma)]\}$$
where
$$t_{v_1 v_2}^{(00)} = t_{v_1 v_2}, \quad t_{v_1 v_2}^{(22)} = t_{v_1 v_2} + 2I_{v_1 v_1 v_2 v_1},$$
$$t_{v_1 v_2}^{(02)} = t_{v_1 v_2}^{(20)} = t_{v_1 v_2} + I_{v_1 v_1 v_2 v_1}.$$
The integrals $I$ should be calculated for the orthogonalized wavefunctions, so that they contain the contributions from non-orthogonality integrals [6]. It should be noted that the dependence of the effective transfer integrals on the atomic ME terms may be rather complicated if we use at solving the atomic problem more advanced approaches [5]. For example, the general Hartree-Fock approximation in the atom theory yields the radial one-electron wave functions which depend explicitly on atomic term. In some variational approaches of the many-electron atom theory the ME wavefunction is not factorized into one-electron ones.

In the works by Shubin and Vonsovsky a quasiclassical approximation was developed [1,7]. This replaces $X$-operators by $c$-numbers which determine the probability amplitudes for the states $|\Gamma\rangle$:
$$X_i(+,0) \to \varphi_i^* \Psi_i, \quad X_i(2,-) \to \Phi_i^* \psi_i, \quad X_i(2,0) \to \Phi_i^* \Psi_i$$
with the subsidiary condition $|\varphi_i|^2 + |\psi_i|^2 + |\Phi_i|^2 + |\Psi_i|^2 = 1$. Such a procedure corresponds to a variational principle with the trial function
$$\varphi = \prod_i (\phi_i^* X_i(+,0) + \psi_i^* X_i(-,0) + \Phi_i^* X_i(2,0) + \Psi_i^*)|0\rangle.$$
This function mixes the Fermi- and Bose-type excitations and thereby does not satisfy the Pauli principle. Nevertheless, the quasiclassical approximation provides a rough description of metal–insulator transition in spirit of the Gutzwiller approximation (see [7]).

The interest in the Hubbard model has been greatly revived after the discovery of high-temperature superconductivity. In particular, the electron states in $CuO_2$-planes of copper-oxide perovskites may be described by the so called Emery Hamiltonian
$$H = \sum_{\mathbf{k}\sigma}\left[\varepsilon p_{\mathbf{k}\sigma}^\dagger p_{\mathbf{k}\sigma} + \Delta d_{\mathbf{k}\sigma}^\dagger d_{\mathbf{k}\sigma} + V_\mathbf{k}(p_{\mathbf{k}\sigma}^\dagger d_{\mathbf{k}\sigma} + d_{\mathbf{k}\sigma}^\dagger p_{\mathbf{k}\sigma})\right] + U\sum_i d_{i\uparrow}^\dagger d_{i\uparrow} d_{i\downarrow}^\dagger d_{i\downarrow}$$
where $\varepsilon$ and $\Delta$ are positions of $p$- and $d$-levels for Cu and O ions, and the $\mathbf{k}$-dependence of matrix elements of $p$–$d$ hybridization for the square lattice is given by
$$V_\mathbf{k} = 2V_{pd}(\sin^2 k_x + \sin^2 k_y)^{1/2}.$$

At $|V_{pd}| \ll \varepsilon - \Delta$ the Emery model is reduced by a canonical transformation to the Hubbard model with strong Coulomb repulsion and the effective Cu–Cu transfer integrals

$$t_{\text{eff}} = V_{pd}^2/(\varepsilon - \Delta).$$

In connection with the high-temperature superconductor theory Anderson [8] put forward the idea of separating spin and charge degrees of freedom in two-dimensional systems by using the representation of slave Bose and Fermi operators

$$c_{i\sigma}^\dagger = X_i(\sigma, 0) + \sigma X_i(2, -\sigma) = s_{i\sigma}^\dagger e_i + \sigma d_i^\dagger s_{i-\sigma}$$

where $s_{i\sigma}^\dagger$ are creation operators for neutral fermions (spinons) and $e_i^\dagger$, $d_i^\dagger$ for charged spinless bosons (holons). The requirement of the Fermi commutation relations for electron operators yields

$$e_i^\dagger e_i + d_i^\dagger d_i + \sum_\sigma s_{i\sigma}^\dagger s_{i\sigma} = 1.$$

The physical sense of such excitations may be explained as follows. Consider the lattice with one electron per site with strong Hubbard repulsion, so that each site is neutral. In the ground resonance valence bond (RVB) state each site takes part in one bond. When a bond becomes broken, two uncoupled sites occur which possess spins of 1/2. The corresponding excitations (spinons) are uncharged. On the other hand, the empty site (hole) in the system carries the charge, but not spin. On the other hand, the empty site (hole) in the system carries the charge, but not spin.

In the half-filled case only spinon excitations with the kinetic energy of order of Heisemberg exchange $|J|$ are present. At doping the system by holes, there occur the current carriers which are described by the holon operators $e_i^\dagger$. In the simplest gapless version, the Hamiltonian of the system for a square lattice may be presented as

$$H = \sum_{\mathbf{k}}[2t(\cos k_x + \cos k_y) - \zeta]e_{\mathbf{k}}^\dagger e_{\mathbf{k}} + 2\sum_{\mathbf{k}}(\Delta + t\delta)(\cos k_x + \cos k_y)(s_{\mathbf{k}\sigma}^\dagger s_{-\mathbf{k}-\sigma}^\dagger + s_{\mathbf{k}\sigma}s_{-\mathbf{k}-\sigma}) + \ldots,$$

with $\zeta$ being the chemical potential, $\Delta$ the RVB order parameter determined by anomalous averages of the spinon operators, $\delta = \langle e^\dagger e \rangle$ the hole concentration. Thus a spin-liquid state can arise (even in purely spin systems without conduction electrons) with long-range magnetic order suppressed, a small energy scale $J$, and a large linear term in specific heat, which is owing to existence of the spinon Fermi surface.

Later, more complicated versions of the RVB theory were developed which use topological consideration and analogies with the fractional quantum Hall effect. These ideas led to rather unusual and beautiful results. For example, it was shown that spinons may obey fractional statistics, i.e. the wavefunction of the system acquires a complex factor at permutation of two quasiparticles.

Taking into account a concrete physical problem, various representation of Hubbard's operators can be used. In the paper [9] a representation of four bosons $p_{i\sigma}$, $e_i$, $d_i$ was proposed which project onto the states $\Gamma = \sigma, 2$ and $0$. Then the Hubbard Hamiltonian takes the form

$$H = \sum_{ij\sigma} t_{ij} f_{i\sigma}^\dagger f_{j\sigma} z_{i\sigma}^\dagger z_{j\sigma} + U\sum_i d_i^\dagger d_i, \quad z_{i\sigma} = e_i^\dagger p_{i\sigma} + p_{i-\sigma}^\dagger d_i$$

with the restrictions

$$\sum_\sigma p_{i\sigma}^\dagger p_{i\sigma} + d_i^\dagger d_i + d_i^\dagger d_i = 1, \quad f_{i\sigma}^\dagger f_{i\sigma} = p_{i\sigma}^\dagger p_{i\sigma} + d_i^\dagger d_i.$$

This representation enables one to reproduce old results on the metal–insulator transition yielding a Gutzwiller-type picture.

Wang [10] proposed the representation with two kinds of Fermi operators $e_i$ and $d_i$ corresponding to holes and doubles:

$$X_i(+, 0) = e_i(1 - d_i^\dagger d_i)(1/2 + s_i^z), \quad X_i(-, 0) = e_i(1 - d_i^\dagger d_i)s_i^-,$$

$$X(2, -) = d_i^\dagger(1 - d_i^\dagger d_i)s_i^+, \quad X(2, +) = d_i^\dagger(1 - d_i^\dagger d_i)(1/2 + s_i^z).$$

The physical spin operators are connected with the pseudospin operators $s_i^\alpha$ by the relation $\mathbf{S}_i = \mathbf{s}_i(1 - d_i^\dagger d_i - e_i^\dagger e_i)$.

Besides that, supersymmetric representations for the Hubbard operators were developed [11].

At theoretical consideration of highly-correlated compounds, including copper-oxide high-$T_c$ superconductors, the $t$–$J$ model (the Hubbard model with $U \to \infty$ and Heisenberg exchange included) is widely exploited. Its Hamiltonian in many-electron representation reads

$$H = -\sum_{ij\sigma} t_{ij} X_i(0\sigma) X_j(\sigma 0) +$$

$$+ \sum_{ij} J_{ij} \left\{ X_i(+-)X_j(-+) + \frac{1}{4}[X_i(++) - X_i(--)][X_j(++) - X_j(--)] \right\}.$$

At derivation of the $t$–$J$ model from the large-$U$ Hubbard model, $J = -4t^2/U$ is the antiferromagnetic kinetic exchange integral. Using the above Fermi-type holon representation, in the case of hole conductivity ($N_e < N$) we get

$$H = -\sum_{ij} t_{ij} e_i^\dagger e_j (1/2 + \mathbf{s}_i \mathbf{s}_j) + \sum_{ij} J_{ij}(1 - e_i^\dagger e_i)(\mathbf{s}_i \mathbf{s}_j - 1/4)(1 - e_j^\dagger e_j).$$

This representation was applied to the magnetic polaron problem in an antiferromagnet.

In the ME representation one can demonstrate that the $t$–$J$ model is a particular case of the narrow-band $s$–$d$ exchange model, corresponding to $I \to -\infty$, $S = 1/2$, $t_\mathbf{k}$ being replaced by $2t_\mathbf{k}$ (the factor of 2 occurs because of equivalence of both electrons in the Hubbard model).

### $s$–$d(f)$ exchange model and related models

The $s$–$d(f)$ exchange model treats the situation where the subsystems of current carriers and magnetic moments are separated. Its Hamiltonian in the simplest case reads

$$H = \sum_{\mathbf{k}\sigma} t_\mathbf{k} c_{\mathbf{k}\sigma}^\dagger c_{\mathbf{k}\sigma} + \sum_\mathbf{q} J_\mathbf{q} \mathbf{S}_{-\mathbf{q}} \mathbf{S}_\mathbf{q} - I \sum_{i\sigma\sigma'} (\mathbf{S}\boldsymbol{\sigma}_{i\sigma\sigma'}) c_{i\sigma}^\dagger c_{i\sigma'}$$

where $\boldsymbol{\sigma}$ are the Pauli matrices, $I$ is the $s$–$d(f)$ exchange parameter.

Being first proposed to describe transport properties of transition $d$-metals [2,12], this model turned out to be very successful to treat the properties of various $d$- and $f$-systems. Recently, the narrow-band $s$–$d$ exchange model with large $|I|$ has been applied to colossal magnetoresistance manganites (the double-exchange problem).

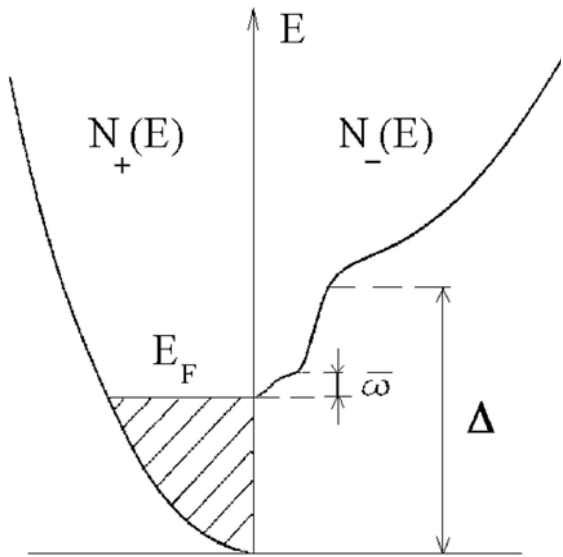

The $s$–$d(f)$ exchange model describes also correlation effects in the half-metallic ferromagnets [13]. These materials have an unusual electronic structure: the states with only one spin projection are present at the Fermi level $E_F$. Thus an important role belongs to the so-called non-quasiparticle (NQP, incoherent) states which arise in the minority- (majority-) spin gap above (below) the Fermi level owing to the electron–magnon interaction. With increasing temperature, the capacity of the spin-polaron tail rapidly increases.

Fig.1. *Density of states in the s–d model with I > 0. At T = 0 the spin-polaron tail of NQP spin-down states reaches $E_F$ (in the case I < 0 NQP spin-up states occur below $E_F$)*

The NQP states make considerable contributions to the electronic properties and can be probed, in particular, by spin-polarized scanning tunneling microscopy (STM). They also lead to observable effects in core–hole spectroscopy, nuclear magnetic relaxation and temperature dependence of impurity resistivity, etc. [13].

Similar (and even more strong) NQP effects occur in a Hubbard ferromagnet with large $U$ and small concentration of doubles $c$. The corresponding Green's function of spin-up electrons has a non-pole form

$$G_{\mathbf{k}\uparrow}(E) = \left\{ E - t_{\mathbf{k}} + (1-c)\left[ \sum_{\mathbf{q}} \frac{n_{\mathbf{k+q}}}{E - t_{\mathbf{k}} + \omega_{\mathbf{q}}} \right]^{-1} \right\}^{-1},$$

$\omega_{\mathbf{q}}$ being the magnon frequency, $n_{\mathbf{k}}$ the Fermi function. The current carriers below the Fermi surface turn out to be fully depolarized. This result has a simple physical meaning. Since the current carriers are spinless doubles (doubly occupied sites), the electrons with spins up and down may be picked up with an equal probability from the states below the Fermi level of doubles. On the other hand, according to the Pauli principle, only the spin down electrons may be added in the singly occupied states in the saturated ferromagnet. Such a picture corresponds to the factorization of the Hubbard operators, $X_i(-,2) \to X_i(-,+)X_i(+,2)$, so that the motion of spin-up electron is incoherent, a bound state of a magnon and spin-down electron being formed.

The filling of the gap by incoherent states is very important for possible applications of half-metallic ferromagnets and other strong metallic magnets in spintronics: these materials really have some advantages only provided that $T \ll T_C$. Since a single-particle Stoner-like theory leads to a much less restrictive (but unfortunately completely wrong) inequality $T \ll \Delta = 2|IS|$, a many-body treatment of the spin polarization problem (inclusion of collective spin-wave excitations) is required.

In many $d$- and $f$-systems the $s$–$d(f)$ exchange interaction has in fact a hybridization nature. Thus the $s$–$d(f)$ model is closely related to the Anderson model. The latter describes the formation of magnetic moments in the situation where highly correlated $d(f)$-electrons do not participate directly in the band motion, but are hybridized with the conduction band states. At neglecting orbital degeneracy the Hamiltonian of periodic Anderson model has the form

$$H = \sum_{\mathbf{k}\sigma} \left[ t_{\mathbf{k}} c^{\dagger}_{\mathbf{k}\sigma} c_{\mathbf{k}\sigma} + \Delta f^{\dagger}_{\mathbf{k}\sigma} f_{\mathbf{k}\sigma} + V(c^{\dagger}_{\mathbf{k}\sigma} f_{\mathbf{k}\sigma} + f^{\dagger}_{\mathbf{k}\sigma} c_{\mathbf{k}\sigma}) \right] + U \sum_{i} f^{\dagger}_{i\uparrow} f_{i\uparrow} f^{\dagger}_{i\downarrow} f_{i\downarrow}.$$

A number of rare-earth elements (Ce, Sm, Eu, Tm, Yb) do not possess a stable valence, but vary it in different compounds. In some systems these elements may produce so-called mixed (or intermediate) valent state which is characterized by non-integer number of $f$-electrons per atom. Such a situation may occur provided that the configurations $4f^n(5d6s)^m$ and $4f^{n-1}(5d6s)^{m+1}$ are nearly degenerate, so that inter-configuration fluctuations are strong. In metallic systems, this corresponds to the $f$-level located near the Fermi energy, $f$-states being hybridized with conduction band states.

The intermediate valence (IV) state is characterized by the single line in Moessbauer experiments, which has an intermediate position. On the other hand, in "fast" X-ray experiments two lines are seen, which correspond to the configurations $f^n$ and $f^{n-1}$. A peculiar feature of the transition into IV state is also the change of the lattice parameter to a value which is intermediate between those for corresponding integer-valent states. Besides that, IV compounds possess at low temperatures substantially enhanced electronic specific heat and magnetic susceptibility.

In a sense, IV systems may be treated as high-$T_K$ Kondo lattices which are considered below. Unlike the Kondo lattice state, not only spin, but also charge fluctuations play an important role in the IV state. The intermediate valent situation corresponds to the situation where the width of $f$-peak owing to hybridization, $\Gamma = \pi V^2 N(E_F)$, is small in comparison with the distance $|\Delta| = |\varepsilon_f - E_F|$. The simplest theoretical model for description of the IV state is the spinless

Falicov–Kimball model

$$H = \sum_{\mathbf{k}}\left[t_{\mathbf{k}}c_{\mathbf{k}}^{\dagger}c_{\mathbf{k}} + \Delta f_{\mathbf{k}}^{\dagger}f_{\mathbf{k}} + V(c_{\mathbf{k}}^{\dagger}f_{\mathbf{k}} + f_{\mathbf{k}}^{\dagger}c_{\mathbf{k}})\right] + G\sum_{i}f_{i}^{\dagger}f_{i}c_{i}^{\dagger}c_{i}$$

where $G$ is the parameter of on-site $d$–$f$ Coulomb repulsion. This Hamiltonian enables one to take simply into account strong on-site $f$–$f$ repulsion (in the spinless model, doubly-occupied states are forbidden by the Pauli principle) and is convenient at description of valence phase transitions, the interaction $G$ being important for many-electron "exciton" effects. The Falicov–Kimball model may be generalized by inclusion of Coulomb interaction at different sites, which permits to describe charge ordering.

**The Kondo effect**

Due to great importance of the Kondo effect, $s$–$d(f)$ exchange model is often (although not quite correctly) called the Kondo-lattice model. This effect was first discussed in connection with the problem of resistivity minimum observed in diluted alloys of transition metals. Kondo demonstrated that in the third order of perturbation theory the $s$–$d$ exchange interaction results in a singular $\ln T$ - correction to resistivity owing to many-body effects (Fermi statistics). It turns out that at low temperatures a peculiar Kondo state is formed. This can be described in terms of a narrow many-particle Abrikosov–Suhl resonance at the Fermi level with a width of order of the Kondo temperature $T_K$. The new Fermi-liquid state is characterized by large many-electron renormalizations, $T_K$ playing the role of the effective degeneracy temperature. Although bare $s$–$d(f)$ parameter is small, the effective (renormalized) $s$–$d(f)$ coupling becomes infinitely strong, so that the impurity magnetic moment is totally compensated (screened) by conduction electrons.

To describe the formation of the singlet Kondo state in the strong coupling region we may use the Hamiltonian of the SU(N) Anderson-lattice model

$$H = \sum_{\mathbf{k}m}t_{\mathbf{k}}c_{\mathbf{k}m}^{\dagger}c_{\mathbf{k}m} + \Delta\sum_{im}X_i(mm) + V\sum_{\mathbf{k}m}[c_{\mathbf{k}m}^{\dagger}X_{\mathbf{k}}(0m) + X_{-\mathbf{k}}(m0)c_{\mathbf{k}m}]$$

($m = 1, \ldots, N$). This model is convenient at describing the inter-configuration transitions $f^0$–$f^1$ (cerium, $J = 5/2$) or $f^{14}$–$f^{13}$ (ytterbium, $J = 7/2$) and is treated often within the $1/N$-expansion. It may be mapped by a canonical transformation, which excludes the hybridization, onto the Coqblin–Schrieffer model

$$H_{CS} = \sum_{\mathbf{k}m}t_{\mathbf{k}}c_{\mathbf{k}m}^{\dagger}c_{\mathbf{k}m} - I\sum_{imm'}X_i(mm')c_{\mathbf{k}m'}^{\dagger}c_{\mathbf{k}m}, \quad I = V^2/\Delta.$$

To avoid difficulties owing to complicated commutation relations for the $X$-operators, the representation may be used

$$X_i(m0) = f_{im}^{\dagger}b_i^{\dagger}, \quad X_i(m'm) = f_{im'}^{\dagger}f_{im}, \quad X_i(00) = b_i^{\dagger}b_i$$

where $f^{\dagger}$ are the Fermi operators, $b^{\dagger}$ are the auxiliary (slave) Bose operators, which satisfy

$$\sum_m X_i(mm) + X_i(00) = \sum_m f_{im}^{\dagger}f_{im} + b_i^{\dagger}b_i = 1.$$

The parameter $\langle b_i \rangle$ renormalizes the hybridization matrix elements. The SU(N) Anderson model permits a description of the crossover to the coherent regime in the Kondo lattice [14]. The temperature dependence of effective hybridization parameter can be obtained in the form

$$V_{\text{eff}}^2 \sim \langle b_i^{\dagger}b_i\rangle \sim \varphi(T), \quad \varphi(T) = (N + e^{-T_K/T} + 1)^{-1} = \begin{cases} 1, & T \ll T_{\text{coh}}, \\ O(1/N), & T_{\text{coh}} \ll T \ll T_K \end{cases}$$

with the coherence temperature $T_{\text{coh}} = T_K/\ln N$.

The Kondo-lattice state may be considered as the nearly integral limit of the IV state (the valence change does not exceed of a few percents). At the same time, delocalization of $f$-states can be obtained directly in the *s–d(f)* exchange model. A special mean-field approximation for the ground state of magnetic Kondo lattices [15] exploits the Abrikosov pseudofermion representation for spin operators $S = 1/2$ and reduces the *s–f* exchange model to an effective hybridization model. The corresponding energy spectrum contains narrow density-of-states peaks owing to the pseudofermion contribution. Thus *f*-pseudofermions become itinerant in the situation under consideration. This fact, although being not obviously understandable, is confirmed by observation of large electron mass in de Haas – van Alphen experiments.

Starting from the middle of 1980s [16,17], anomalous rare-earth and actinide compounds are extensively studied. They include Kondo lattices (with moderately enhanced electronic specific heat) and heavy-fermion systems demonstrating a huge linear specific heat. Main role in the physics of the Kondo lattices belongs to the interplay of the on-site Kondo screening and intersite exchange interactions. Following to the Doniach criterion [18], it was believed in early works that the total suppression of either magnetic moments or the Kondo anomalies takes place. However, more recent experimental data and theoretical investigations made clear that the Kondo lattices as a rule demonstrate magnetic ordering or are close to this. This concept was consistently formulated and justified in the papers [19] treating the mutual renormalization of two characteristic energy scales: the Kondo temperature $T_K$ and spin-fluctuation frequency $\bar{\omega}$. A simple scaling consideration of this renormalization process in the *s–f* exchange model [19] yields, depending on the values of bare parameters, both the usual states (a non-magnetic Kondo lattice or a magnet with weak Kondo corrections) and the peculiar magnetic Kondo-lattice state. In the latter state, small variations of parameters result in strong changes of the ground-state moment. Thereby high sensitivity of the ground-state moment to external factors like pressure and doping by a small amount of impurities (a characteristic feature of heavy fermion magnets) is naturally explained.

During 1990s, a number of anomalous *f*-systems ($U_xY_{1-x}Pd_3$, $UPt_{3-x}Pd_x$, $CeCu_{6-x}Au_x$, $Ce_7Ni_3$ etc.) demonstrating the non-Fermi-liquid (NFL) behavior have become a subject of great interest (see the review [20]). These systems possess unusual logarithmic or power-law temperature dependences of electron and magnetic properties. The NFL behavior is typical for Kondo systems lying on the boundary of magnetic ordering and demonstrating strong spin fluctuations.

**Summary**

Being formulated already in the first half of XX century, the polar and *s–d(f)* exchange models still work successfully in the solid state physics. They provide a basis for new theoretical concepts describing physical phenomena discovered by experimentators. The model approaches which include effects of strong electron correlations in *d*- and *f*-compounds turn out to be very useful from the point of view of the qualitative microscopic description.

The spectrum of highly-correlated systems is often described in terms of auxiliary (slave) Fermi and Bose operators, which correspond to quasiparticles with exotic properties (neutral fermions, charged bosons etc.). Last time such ideas have been extensively applied in connection with the unusual spectra of high-$T_c$ superconductors and heavy-fermion systems. Investigation of these problems leads to complicated mathematics, which uses the whole variety of modern quantum field theory methods, and very beautiful physics. For example, description of the Fermi-liquid state in terms of Bose excitations becomes possible. These concepts change essentially classical notions of the solid state theory. Modern many-particle physics is intimately connected to other fields of science: nuclear and elementary-particle physics, cosmology, quantum technologies, biology etc.

The work was supported in part by the Program "Quantum Physics of Condensed Matter" from Presidium of Russian Academy of Sciences.